\documentclass{elsart}
\usepackage{natbib}
\usepackage{graphicx}
\usepackage{amssymb}
\usepackage{amsmath}
\usepackage{bm}

\def\url#1{{\ttfamily\def\/{/\discretionary{}{}{}}#1}}
\def\bibcode#1{}

\begin{document}
\begin{frontmatter}
\title{Cosmological science enabled by Planck}
\author{Martin White\thanksref{mjwemail}}
\address{Departments of Physics and Astronomy,\\
University of California, Berkeley, CA 94720}
\thanks[mjwemail]{mwhite@berkeley.edu}

\begin{abstract}
{\sl Planck\/} will be the first mission to map the entire cosmic microwave
background (CMB) sky with mJy sensitivity and resolution better than $10'$.
The science enabled by such a mission spans many areas of astrophysics and
cosmology.  In particular it will lead to a revolution in our understanding
of primary and secondary CMB anisotropies, the constraints on many key
cosmological parameters will be improved by almost an order of magnitude (to
sub-percent levels) and the shape and amplitude of the mass power spectrum
at high redshift will be tightly constrained.
\end{abstract}
\end{frontmatter}

\section{Introduction}

{\sl Planck\/} will be the first mission to map the entire cosmic microwave
background (CMB) sky with mJy sensitivity and resolution better than $10'$
\cite{BlueBook}.
The science enabled by such a mission spans many areas of astrophysics and
cosmology, but in this short proceedings I can focus on only a few.  (Further
discussion of the cosmological science enabled by {\sl Planck\/} was covered
by Lloyd Knox in his talk at this meeting.)   In particular I want to
focus on the dramatic revolution {\sl Planck\/} will represent in the study
of primary CMB anisotropies and the universe at $z=10^3$, with its implications
for low-$z$ studies such as those of dark energy.  I also want to make a
push for a CMB-centric view of structure formation which emphasizes the
exquisite constraints on large-scale structure that we already have from the
CMB at high-$z$.

Before I begin with these science topics, it is important to remind ourselves
how revolutionary {\sl Planck\/} will be.  In addition to wider frequency
coverage (crucial for control of foregrounds) and better sensitivity than
{\sl WMAP\/}, {\sl Planck\/} has the resolution needed to see into the
damping tail of the anisotropy spectrum.  In fact {\sl Planck\/} will be the
first experiment to make an almost cosmic variance limited measurement of the
temperature anisotropy spectrum around the $3^{\rm rd}$ and $4^{\rm th}$
acoustic peaks.

\begin{figure}
\begin{center}
\resizebox{2.7in}{!}{\includegraphics{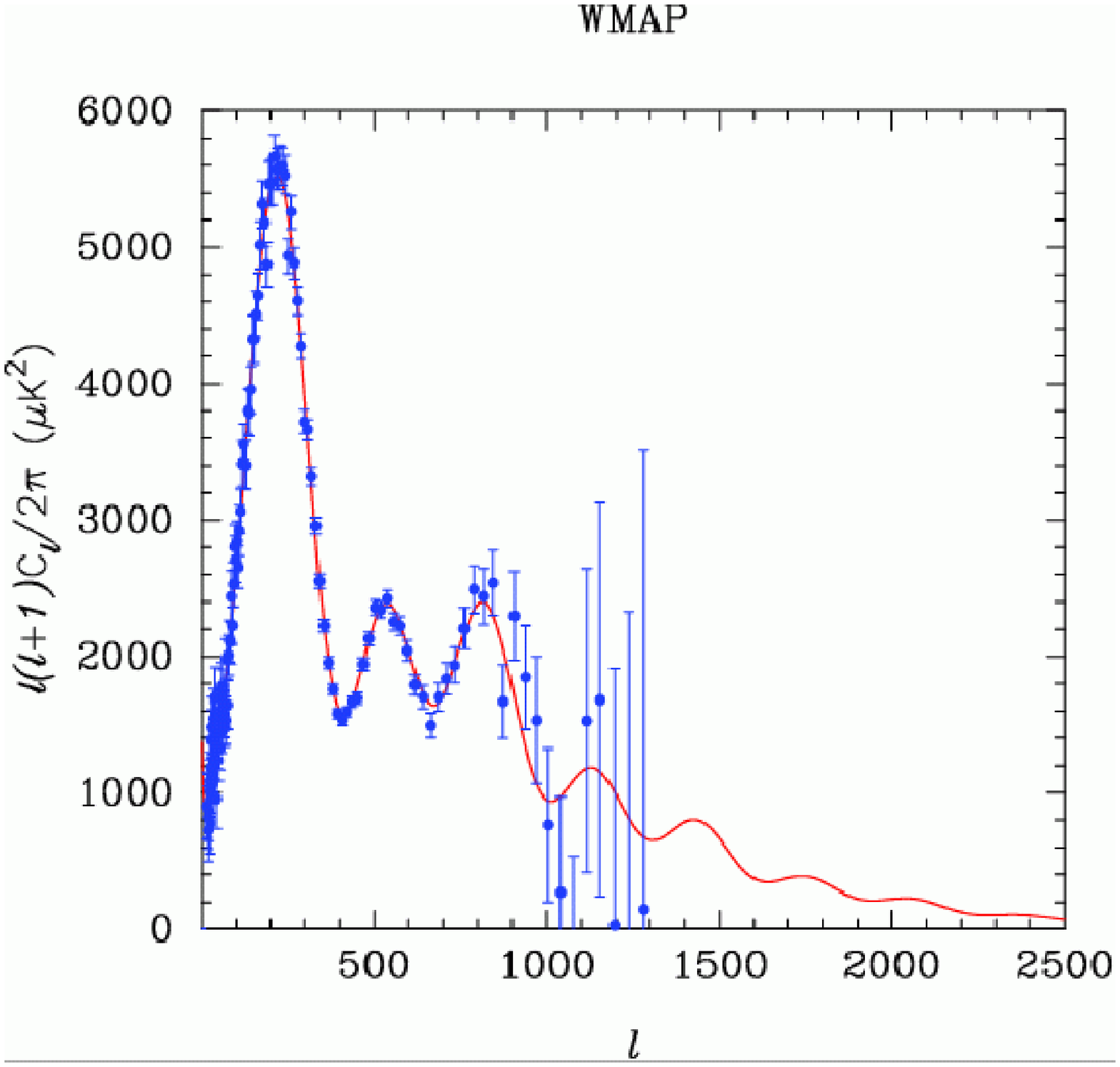}}
\resizebox{2.7in}{!}{\includegraphics{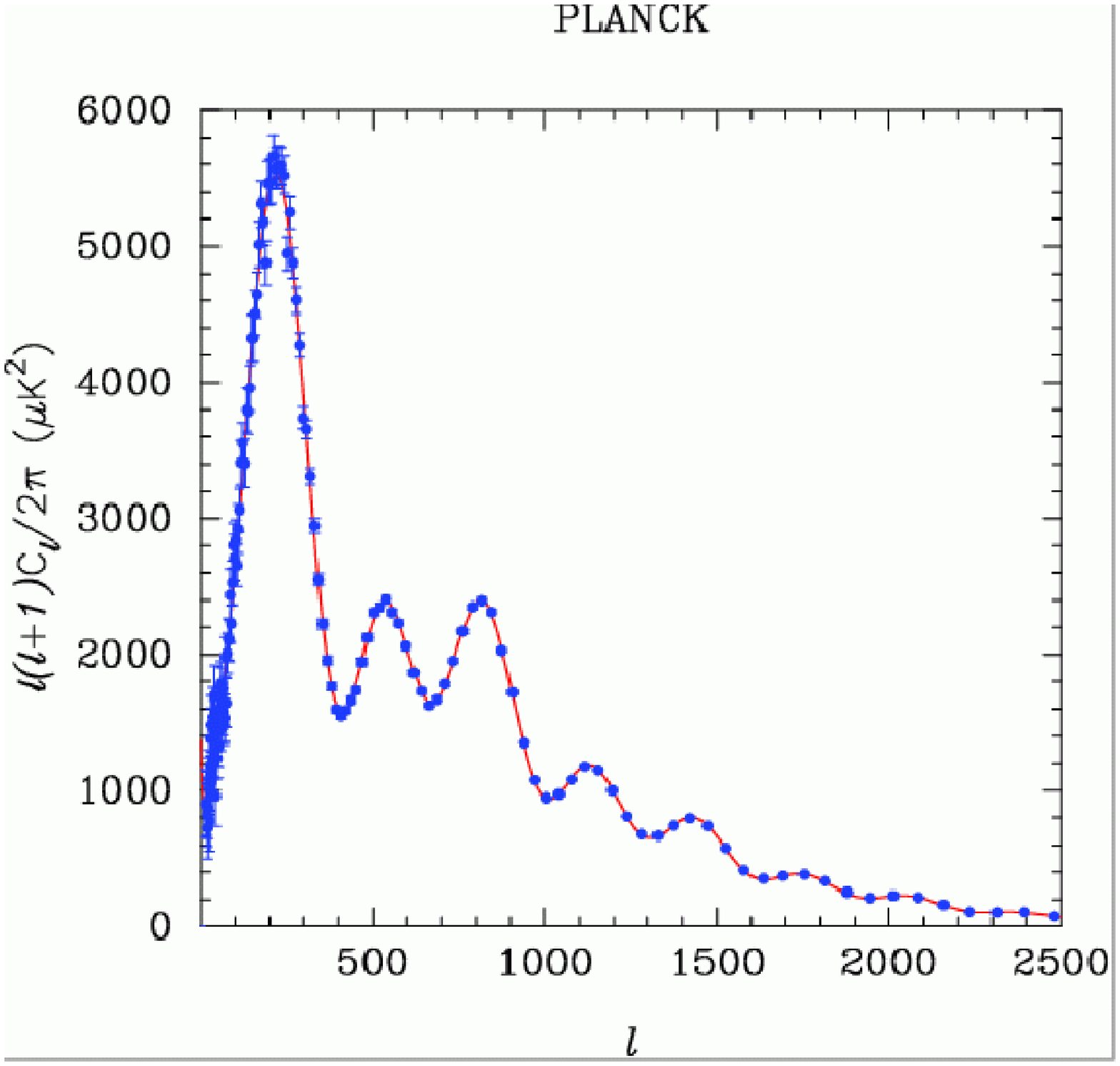}}
\end{center}
\caption{Forecast measurements of the temperature anisotropy power spectrum
{}from 4 years of {\sl WMAP\/} or 1 year of {\sl Planck\/} data assuming
nominal sensitivities.  We have chosen the same binning scheme to show the
advantage that higher resolution and sensitivity confers on {\sl Planck\/}
for high-$\ell$ science.  Figures from \protect\cite{BlueBook}.}
\label{fig:tt}
\end{figure}

\begin{figure}
\begin{center}
\resizebox{2.7in}{!}{\includegraphics{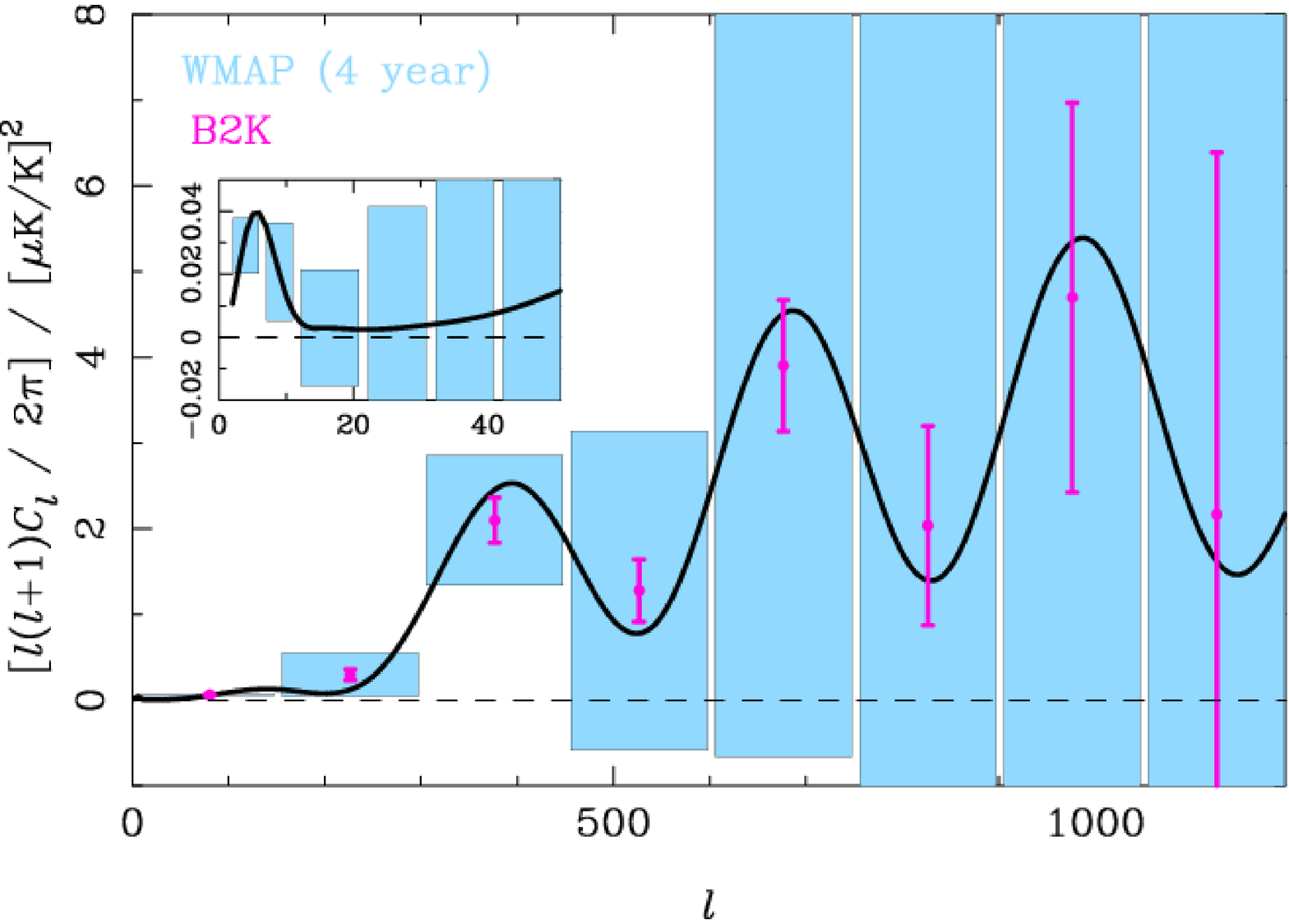}}
\resizebox{2.7in}{!}{\includegraphics{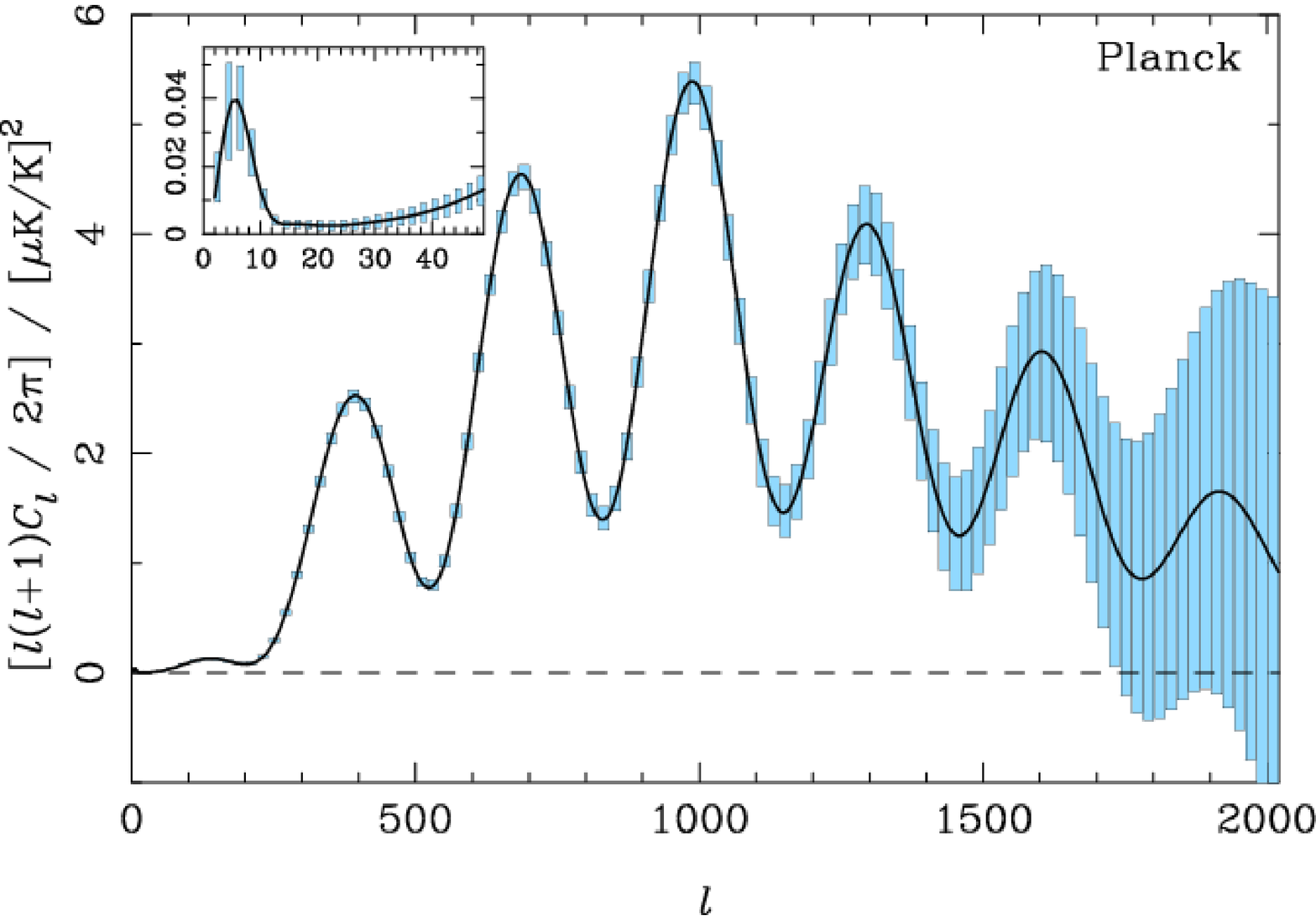}}
\end{center}
\caption{Forecast measurements of the polarization anisotropy power spectrum
{}from 4 years of {\sl WMAP\/} or 1 year of {\sl Planck\/} data assuming
nominal sensitivities.  One can see clearly the advantage that higher
sensitivity confers on {\sl Planck\/} for polarization science.
Figures from \protect\cite{BlueBook}.}
\label{fig:ee}
\end{figure}

What does this dramatic increase in our knowledge of the temperature and
polarization anisotropy spectra tell us about cosmology, fundamental physics
and the formation of structure?  Here I will highlight just a few areas
where we expect a large impact.

\section{The universe at $z=10^3$ and cosmological parameters}
\label{sec:cospar}

\subsection{Constraining cosmological parameters}

It is well known that detailed observations of CMB anisotropy, coupled
with accurate theoretical predictions, constrain the high redshift universe.
The most strongly constrained is the physics which gives rise to the
acoustic peaks in the CMB power spectrum, since that is both where the
measurements are most accurate and the structure the most rich.  To a
first approximation the acoustic peaks constrain the physical matter
density, $\omega_m\equiv\Omega_mh^2$, the physical baryon density,
$\omega_b\equiv\Omega_bh^2$, and an acoustic scale, $\theta_A$ or $\ell_A$
\cite{GrandDad}.
{}From this we can derive other constraints, for example the distance to
last-scattering $D(z=10^3)$.  Currently we know $D(z=10^3)$ to about 2\%
\cite{Spe06}, with the main source of uncertainty coming not from our
knowledge of the peak positions but from the 8\% uncertainty in $\omega_m$.
This translates into an uncertainty in the expansion rate of the universe
near last scattering and hence the distance.

The key to improving our knowledge of $\omega_m$ is the higher peaks.
{\sl Planck\/} should determine $\omega_m$ to $0.9\%$ \cite{BlueBook},
almost an order of magnitude improvement over current knowledge.
In principle this allows us to determine $D(z=10^3)$ to $0.2\%$!
This provides an important constraint for cosmological models, e.g.~on the
dark energy and allows us to calibrate the baryon acoustic oscillation method
for measuring $d_A(z)$ and $H(z)$ in the range $z=0.3-3$ \cite{EisReview}.

\begin{figure}
\begin{center}
\resizebox{5in}{!}{\includegraphics{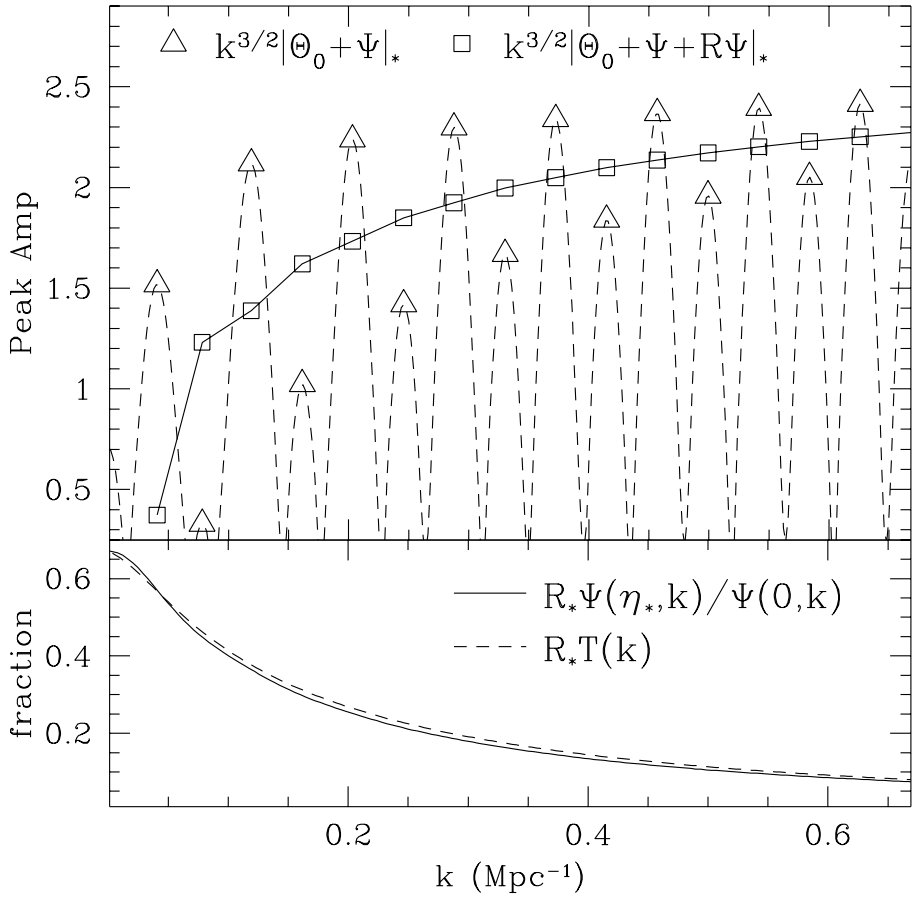}}
\end{center}
\caption{The temperature anisotropy spectrum with the effects of Silk
damping removed.  Concentrating on the triangles in the upper panel we
see the effects of baryon loading in the modulation of the peaks.  If
we remove the modulation the boost at high $\ell$ due to potential decay
becomes apparent (squares, see text).  The lower panel shows by what
fraction the potential has decayed by the present as a function of wavenumber.
Figure taken from \protect\cite{Damp}.}
\label{fig:damp}
\end{figure}

Why are the higher peaks crucial to constraining the matter density?
To understand this let us consider how the temperature anisotropies are formed.
We know that on small scales the CMB anisotropy spectrum is damped by
photon diffusion \cite{Silk,Damp}.  This process is well understood and
essentially independent of the source of the anisotropies.  If we remove
it we can see the combined effects of the baryon loading and the epoch
of equality (Fig.~\ref{fig:damp}).

The baryons give weight to the photon-baryon fluid.  This makes it easier
to fall into a potential well to become a compression and harder to ``bounce''
out to become a rarefaction.  For adiabatic models the baryon loading thus
enhances the compressions (odd peaks) and weakens the rarefactions (even
peaks) leading to an alternating sequence of peak heights.
Superposed upon this alternation is a general rise in power to small scales
-- usually obscured by the effects of the exponential Silk damping.
The power increase arises because at early times -- when the perturbations
giving rise to the higher peaks are entering the horizon -- the baryon-photon
fluid contributes more to the total energy density of the universe than the
dark matter.  The effects of baryon-photon self-gravity enhance the
fluctuations on small scales as follows \cite{AcousticSignatures}.
Since the fluid has pressure it is hard to compress.  This makes the infall
into the potential wells slower than free-fall, retarding the growth in the
overdensity.  Because the overdensity cannot grow rapidly enough the potential
is forced to decay by the expansion of the universe (see lower panel of
Fig.~\ref{fig:damp}).
The photons are then left in a compressed state with no need to fight against
the potential as they leave -- enhancing the small scale power.
As the universe expands and larger scales enter the horizon the dark matter
potentials become increasingly important and the boost is reduced\footnote{If
we were to ignore the effects of neutrinos near equality and the dark energy
at late times the asymptotic value of the excess power would be a factor of
$25$ compared to the low $\ell$ plateau.  The plateau has an amplitude set by
$-\Phi/3$ \cite{SW}.  The infall into the potential well and subsequent decay
boosts the power by $2\Phi$ making the small-scale effective temperature
perturbation $(2-1/3)\Phi=(5/3)\Phi$.  In reality because of the effects of
dark energy and neutrinos the effect is more like a factor of $15$.}.

Thus measuring the higher peaks constrains the behavior of the potentials,
which respond to the expansion rate of the universe near last scattering.
If dark energy is sub-dominant at high $z$ this becomes a constraint on
the epoch of equality, $z_{\rm eq}$, or the matter density.
Since it is able to make an almost cosmic variance limited measurement of
the higher acoustic peaks, {\sl Planck\/} provides us with an unparalleled
constraint on the (physical) matter density.

\subsection{The CMB and dark energy}

The nature of the dark energy believed to be causing the accelerated
expansion of the universe is one of the most important questions facing
cosmology, with implications for our understanding of physics at the
deepest levels.  For this reason the community has been pursuing dark
energy science with a number of different probes.  It is sometimes
stated that the CMB does not directly constrain dark energy, and this is
true.  However it is important to point out that almost all of the methods
which seek to constrain the dark energy do much better if they include
information from the CMB.  In fact most analyses include {\sl WMAP\/}
(or projected {\sl Planck\/}) priors as a matter of course.  As a field
we have not been particularly effective in promoting the importance of
improved CMB anisotropy/polarization measurements for future dark energy
experiments, so let me take some time to show one example here: baryon
acoustic oscillations.

\subsubsection{Acoustic oscillations and the sound horizon}

\begin{figure}
\begin{center}
\resizebox{2.5in}{!}{\includegraphics{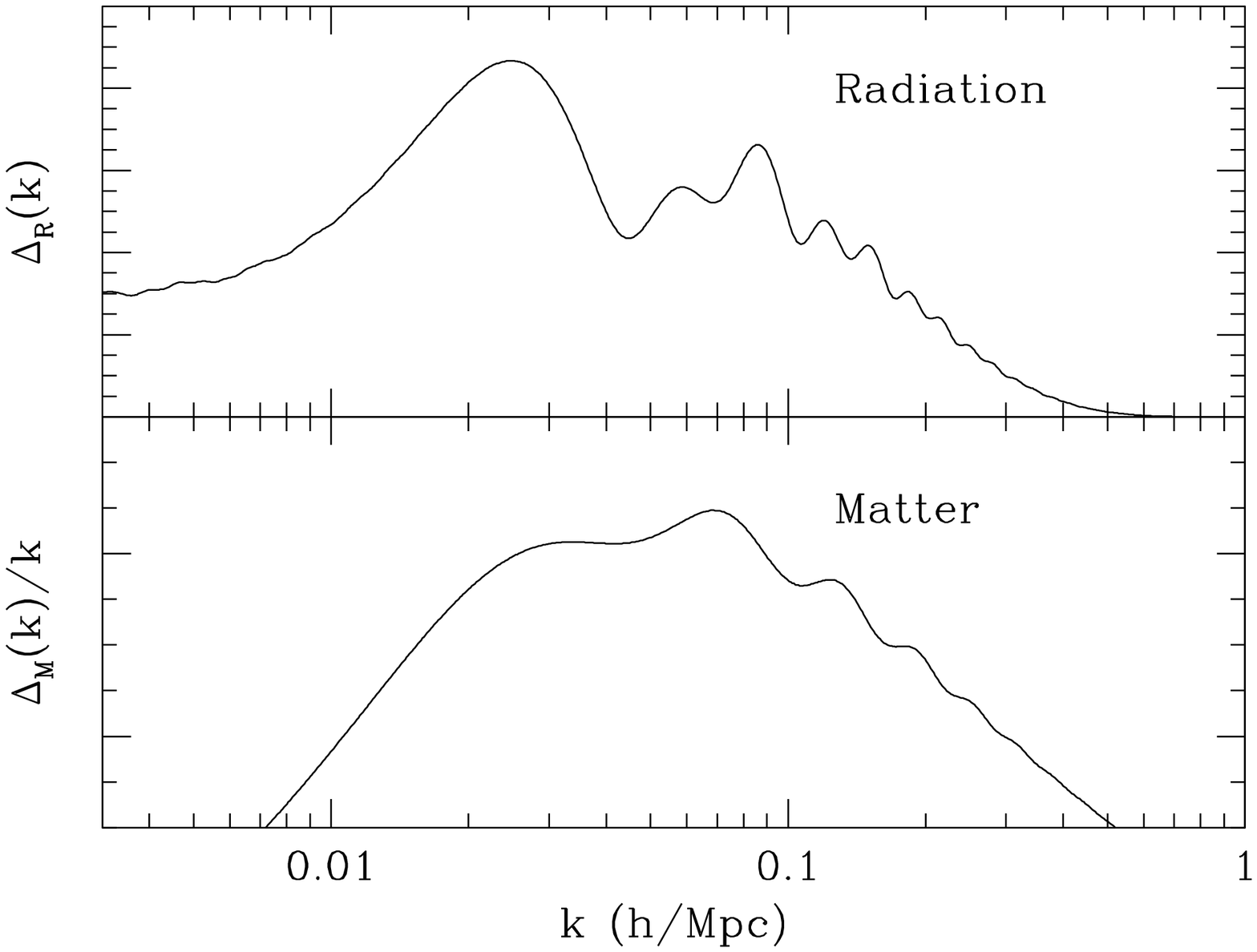}}
\resizebox{2.5in}{!}{\includegraphics{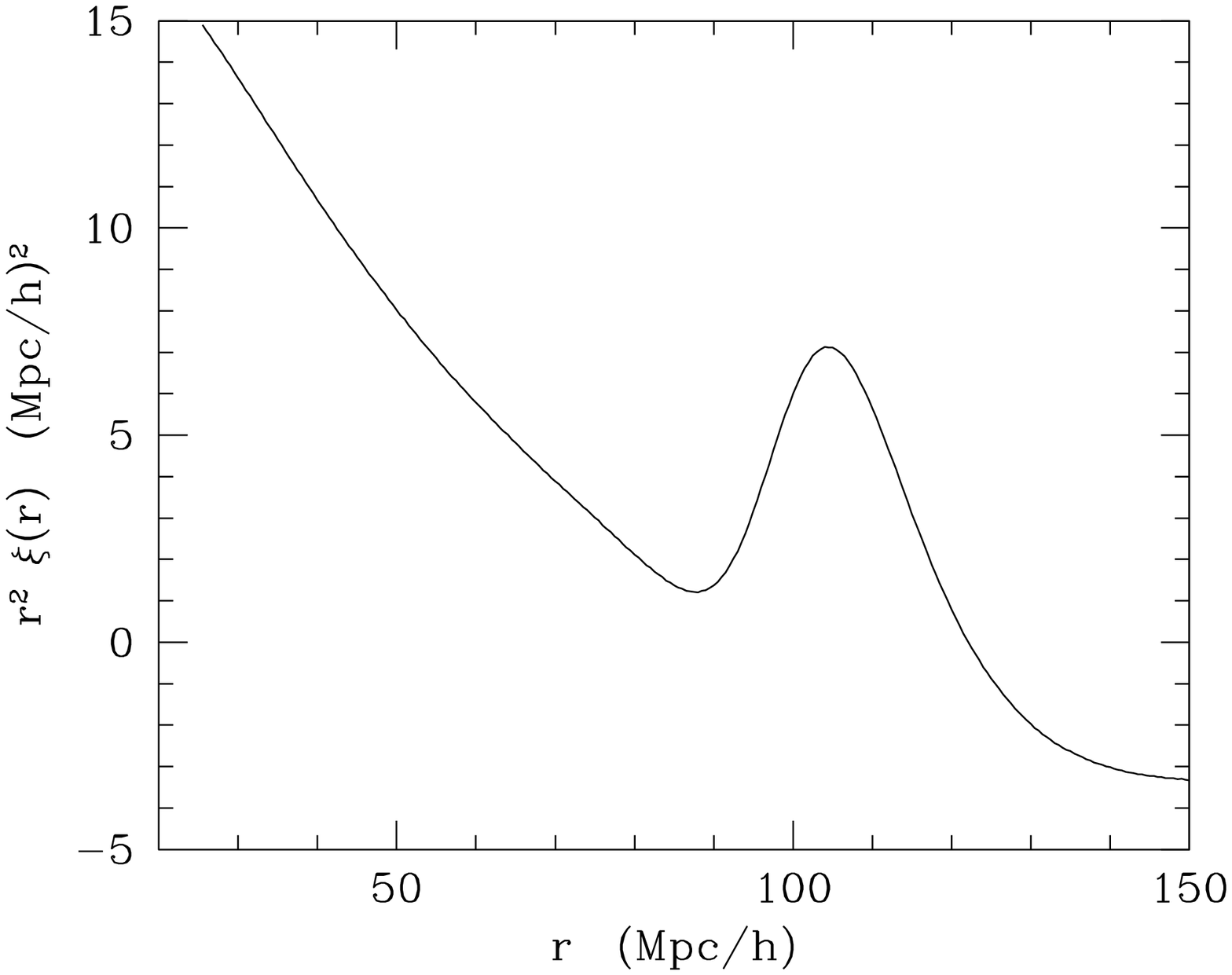}}
\end{center}
\caption{(Left) The linear theory matter and radiation power spectra
vs.~wavenumber.  The upper panel shows the contribution to the RMS
temperature fluctuation per logarithmic interval in wavenumber, and is
closely related to the more familiar angular power spectrum plotted
vs.~angular wave mode $\ell$.
The lower panel shows the (dimensionless) mass power spectrum (divided by
wavenumber $k$).  Note the similar scale of the acoustic oscillations in
each spectrum, and the damping to higher wavenumber.
(Right) The correlation function, or Fourier transform of the power
spectrum plotted in the lower left panel.  Note that the almost harmonic
series of peaks in Fourier space translates into a single well defined
peak in real space with a width of ${\mathcal O}(10\%)$.
From \protect\cite{BaryonPM}.}
\label{fig:rms}
\end{figure}

\begin{figure}
\begin{center}
\resizebox{2.5in}{!}{\includegraphics{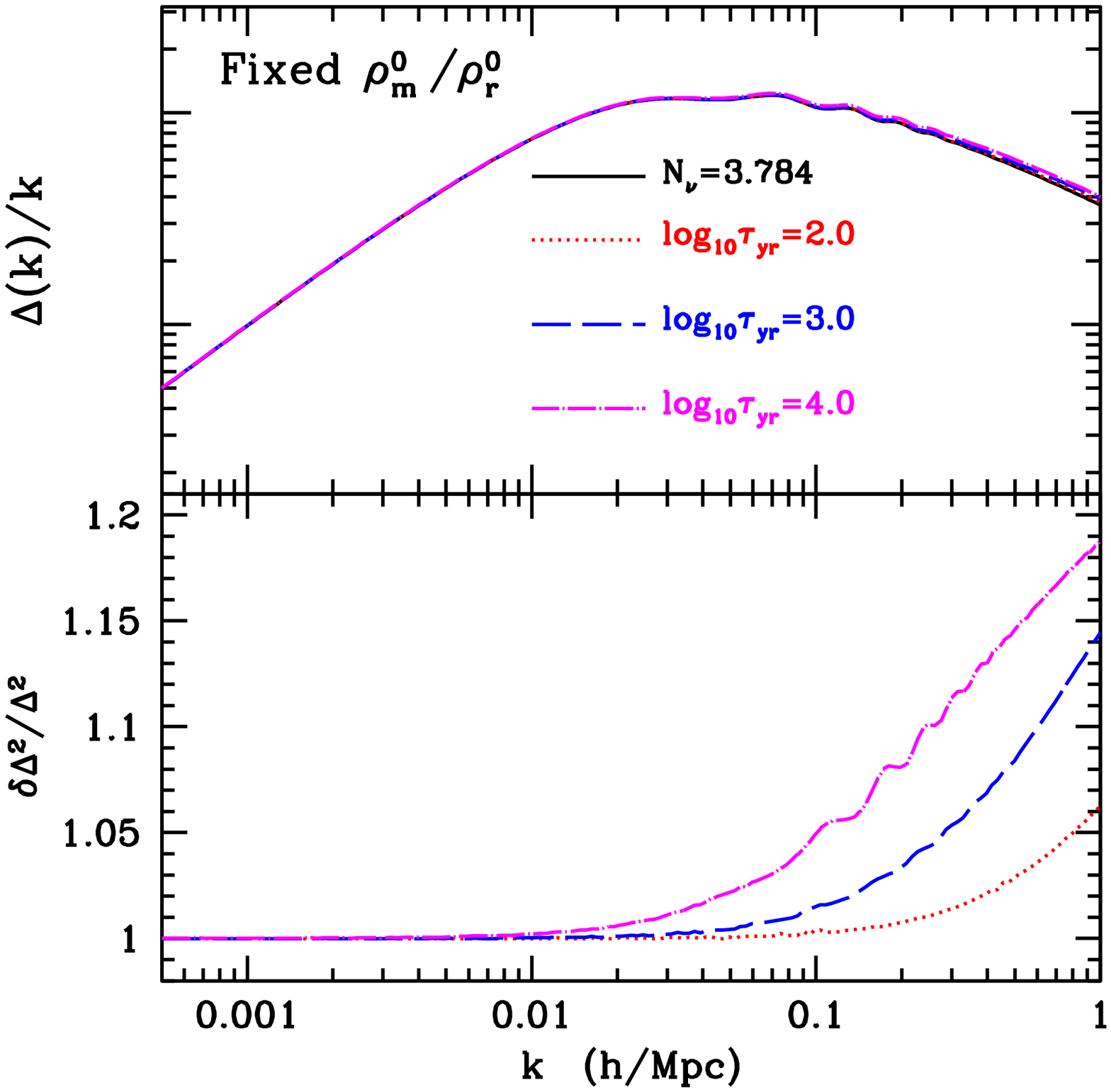}}
\resizebox{2.5in}{!}{\includegraphics{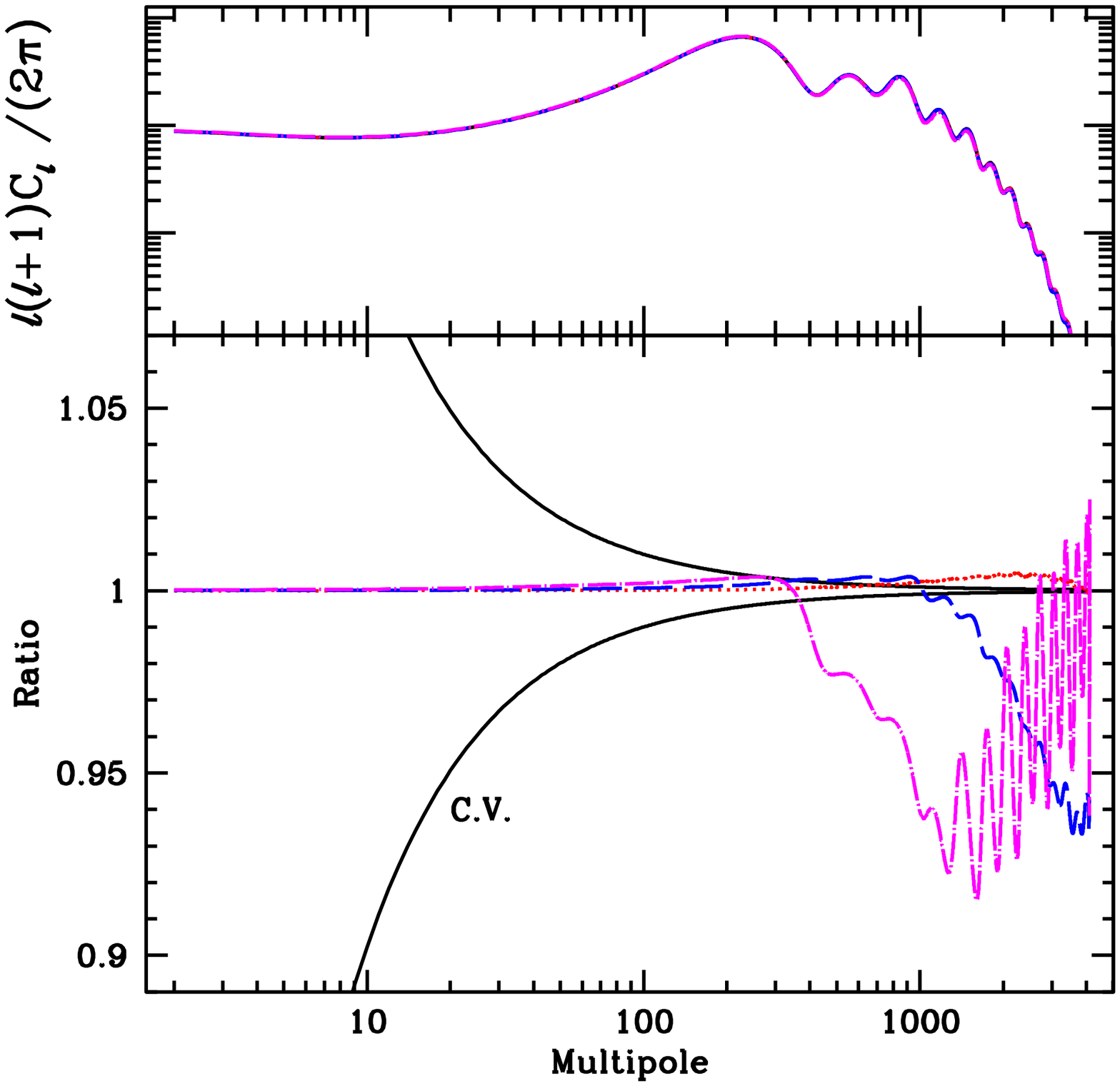}}
\end{center}
\caption{A plot of the power spectra for decaying neutrino scenarios
(see text).  The left upper panel shows the (dimensionless) mass power spectrum
(divided by wavenumber $k$) for models which have $10\%$ more matter and
radiation than the standard model.  The lower left panel shows the ratio of
the curves to the $N_\nu=3.784$ result.  The right panels show the radiation
spectra for the same models, along with cosmic variance error bars averaged
in bins of width $\Delta\ell/\ell=0.1$.}
\label{fig:decneut}
\end{figure}

The idea behind the baryonic acoustic oscillation (BAO) method is to
make measurements of $d_A(z)$ and $H(z)$ using a calibrated standard
ruler which can be measured at a number of redshifts \cite{EisReview}.
The CMB provides the calibrated ruler through its measurement of the
sound horizon:
\begin{equation}
  s \equiv \int_0^{t_{\rm rec}} c_s(1+z)dt
  = \int_{z_{\rm rec}}^\infty \frac{c_s\,dz}{H(z)}
  \quad .
\label{eqn:sdef}
\end{equation}
The sound horizon is extremely well constrained by the structure of the
acoustic peaks.  For example from \cite{Spe06} we find
$s=147.8\pm 2.6\ {\rm Mpc}=\left(4.56\pm0.08\right)\times10^{24}\ {\rm m}$.
As can be seen from Eq.~(\ref{eqn:sdef}) the sound horizon depends on the
expansion history (matter-radiation equality) and the sound speed
(baryon-photon ratio).  Once $s$ is known and the angular scale of the peaks,
$\theta_A$, is measured the distance to last-scattering follows from
$s=D\theta_A$.  The same physical scale is imprinted upon the matter power
spectrum (see Fig.~\ref{fig:rms}), and can serve as a calibrated standard
ruler at lower $z$.

If we Fourier transform the almost harmonic series of peaks seen in
Fig.~\ref{fig:rms} we predict that the correlation function should have a
single well-defined peak at $\sim 100\,h^{-1}$Mpc
(see \cite{ESW} for a discussion of the physics of the BAO in Fourier
and configuration space).
This feature has now been seen by several groups \cite{SDSS} in both
configuration, $\xi(r)$, and Fourier, $\Delta^2(k)$, space at intermediate
redshift, $z\sim 0.35$.  This measurement, along with the CMB, is enough
to show the existence of dark energy but larger surveys are needed to
constrain its properties.

\subsubsection{What could go wrong?}

With so much riding on the CMB calibration it is important to ask what
could go wrong?  Recall the method hinges on the ability to predict $s$,
for which we need $z_{\rm rec}$, $c_s$ and $H(z)$.  It turns out that
recombination is very robust, and our current uncertainties in recombination
\cite{Recomb} lead to shifts in the sound horizon well below a percent.
If we assume the standard radiation content ($3$ nearly massless neutrino
species plus photons) knowing $\rho_\gamma$ from $T_\gamma$ gives $\omega_r$.
Then knowing $z_{\rm eq}$ is the same as knowing $\omega_m$ and $H(z)$.
But what if $\omega_r$ was different?  Could we mistake $\nu$ for DE?

It turns out that as long as $z_{\rm eq}$ is still known well from the CMB
is doesn't matter!  We would misestimate $\omega_m$ however in comparing our
standard ruler at $z\sim 1$ and $z\sim 10^3$ the same $\omega_m$ prefactor
enters $H^{-1}$, $d_A$ and $s$: each scales as $\omega_m^{-1/2}$.  Thus all
distance ratios and DE inferences go through unchanged \cite{EisWhi}.  What
we do is misestimate the overall scale, and hence $H_0$!  It is ironic that
we may end up understanding quantum gravity and the mysterious dark energy
but still be uncertain about the Hubble constant\footnote{I term this the
Hubble uncertainty principle.}.

What about more bizarre histories?  As an example imagine a non-relativistic
particle of mass $m$ which decays with lifetime $\tau$ into massless neutrinos
\cite{DecNeut}.
We arrange $m$ and $\tau$ so that there is 10\% more radiation today than in
the standard model, but increase $\omega_m$ by 10\% so that equality is held
fixed.  Since it is equality that primarily controls the decay of the
potentials at early times the CMB fluctuations look very similar.  However,
because the universe would be slightly more matter dominated at early
times (when the massive particle was a non-negligible contribution to the
total energy density) we would expect excess power on small scales, and we
can shift the acoustic peaks.
Can this lead to a false signature of dark energy?

We show in Fig.~\ref{fig:decneut} the mass and CMB temperature power spectra
for a sequence of models with $\log_{10}\tau_{\rm yr}=2$, 3 and 4.  While
one can see subtle shifts in the sound horizon, any model which appreciably
shifts $s$ changes the temperature anisotropies at high $\ell$ enough to be
easily seen by {\sl Planck}.  Thus while one may not be able to fit the
spectrum with a standard model, one would not mistake strange physics at
$z\sim 10^3$ for dark energy at $z\sim 0$.

\subsection{Conclusions}

To recap the main points of this section, {\sl Planck\/} will dramatically
improve our knowledge of the physical conditions in the universe at
$z\sim 10^3$.  The physical matter and baryon densities and the distance
to last-scattering will be known to sub-percent accuracy.  The epoch of
equality will be tightly constrained, as will extra species, anisotropic
stresses and decaying components at high redshift \cite{EisWhi}.

\section{The CMB prior and structure formation}

Already with {\sl WMAP\/}, and certainly after {\sl Planck\/}, we will
have very precise knowledge of the universe at $z=10^3$.  We will have
tightly constrained the densities of matter and baryons, the amplitude
of the fluctuations in the linear phase over 3 decades in length scale
and the shape of the primordial power spectrum.  {\em Our knowledge of
the physical conditions and large-scale structure at $z=10^3$ will be
better than our knowledge of such quantities at $z=0$.}
One should not ignore this dramatic advance in our knowledge -- when
forecasting the future we should hold the $z=10^3$ universe ``fixed'',
not the $z=0$ one.  This is equivalent to imposing strong CMB priors
on future measurements.

\begin{figure}
\begin{center}
\resizebox{5in}{!}{\includegraphics{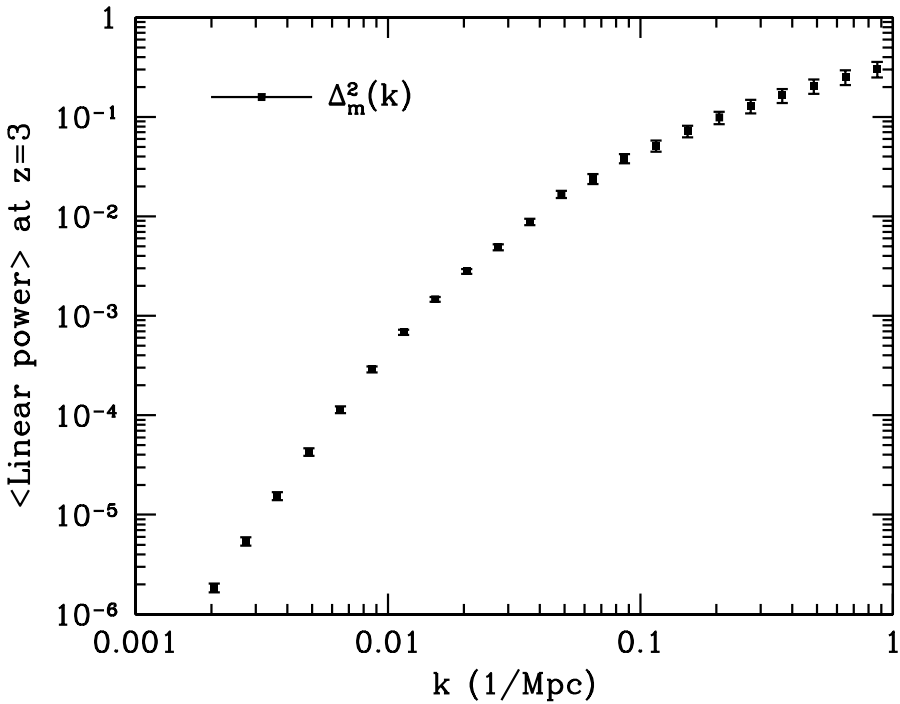}}
\end{center}
\caption{The range of $\Delta_m^2(k)$ allowed by the WMAP 3yr data assuming
a standard CDM model.  The data already constrain
$\Delta^2\left(k\simeq 0.01\,{\rm Mpc}^{-1}\right)$ to 7\%.  This drops to
3\% if the degeneracy with $\tau$ is controlled for.}
\label{fig:matdelk}
\end{figure}

As an example, knowing $\omega_m$ and $\omega_b$ allows us to predict
the shape of the linear theory (matter) power spectrum extremely accurately
over many orders of magnitude in length scale, providing that the lengths
are measured in Mpc (or meters) rather than $h^{-1}$Mpc as would be more
familiar from low-$z$ measurements.  Note that knowing $\omega_m$ and
$z_{\rm eq}$ fixes $H(z)$ at high-$z$
\begin{equation}
  H(z\gg 1)\simeq H_0\sqrt{ \Omega_m(1+z)^3+\Omega_r(1+z)^4 }
  \propto \omega_m^{1/2}\sqrt{ 1 + \frac{1+z}{1+z_{\rm eq}} }
  \quad .
\end{equation}
The amplitude of the fluctuations is well constrained by anisotropy
measurements, up to a degeneracy with $\tau$ --- the constrained
quantity is roughly $\delta_m e^{-\tau}$.  Further, unless dark energy
is important at $z\gg 1$, we can evolve the fluctuations reliably from
$z\sim 10^3$ to lower $z$, since $\delta_m\propto a$ for most of the time.
In combination this enables us to constrain the high-$z$ matter power
spectrum (with lengths measured in {\em physical\/} units).  For example
Fig.~\ref{fig:matdelk} shows the range of matter power spectra at $z=3$
allowed by the {\sl WMAP\/} 3yr data assuming a standard\footnote{We neglect
here a possible running of the spectral index, massive neutrinos or a warm
dark matter candidate.  These will increase the uncertainty on small scales
and may be relevant for the formation of the first structures.}
$\Lambda$CDM model and that dark energy is negligible for $z\ge 3$.
The error bars expand slightly at high-$k$ if we additionally allow massive
neutrinos, but near the scales contributing to the first acoustic peak
($k\simeq 10^{-2}\,{\rm Mpc}^{-1}$) the constraint is already 7\% in power.
Half of the uncertainty comes from the uncertainty in the optical depth,
$\tau$.  If we remove that degeneracy the constraint becomes 3\% in power or
1.5\% in amplitude!  We expect this to improve with future {\sl WMAP\/} data,
but with {\sl Planck\/} the uncertainty will drop to sub-percent levels even
with improved modeling of the reionization epoch.
Thus in a post-{\sl Planck\/} world the uncertainty in large-scale structure
comes from the extrapolation from $z\sim 3$ to $z=0$ (which depends on the
nature of the dark energy) and the conversion between physical distances and
redshift space measures (which depend on $h$).  The former lead to vertical
shifts in the spectrum, while the latter give horizontal shifts.

\section{Conclusions}

{\sl Planck\/} will provide a dramatic advance in our knowledge of
primary and secondary CMB anisotropies.  The constraints on many key
cosmological parameters will be dropped to percent, or sub-percent,
levels and the shape and amplitude of the mass power spectrum at
high redshift will be tightly constrained.  Beyond our desire to know
the basic parameters of the universe accurately, and to perform truly
precision tests of our cosmological model, the increase in precision
will be important for a host of low redshift experiments, including
those that aim to constrain the nature of the dark energy.

I would like to thank the organizers of this conference for a pleasant
and productive meeting, and Daniel Eisenstein and Wayne Hu for conversations
and collaborations upon which some of this work rests.  I am grateful to the
many members of the {\sl Planck\/} collaboration who have labored tirelessly
to make {\sl Planck\/} a reality, and especially to Charles Lawrence for
his leadership and tireless enthusiasm -- and the chocolate donuts.
MJW was supported in part by NASA.

\end{document}